\documentclass[12pt,aps,prb,preprint]{revtex4}   

\usepackage{amsmath}    
\usepackage{graphicx}   

\begin{document}

\title{Non-equilibrium statistical mechanics:  a solvable model}
\author{I. Mazilu}
\email{mazilui@wlu.edu}
\author{H. T. Williams}
\email{williamsh@wlu.edu}
\affiliation{Department of Physics and Engineering \\ Washington and Lee University, Lexington, VA  24450, USA}
\date{\today}

\graphicspath{%
    {converted_graphics/}
    {C:/}
    {./}
}
\begin{abstract}
  A two-temperature linear spin model is presented that allows an easily understandable introduction to non-equilibrium statistical physics.  The model is one that includes the concepts that are typical of more realistic non-equilibrium models but that allows straightforward steady state solutions and, for small systems, development of the full time dependence for configuration probabilities.  The model is easily accessible to upper-level undergraduate students, and also provides a good check for computer models of larger systems.
\end{abstract}
\maketitle

\section{Introduction}

For over a century, the statistical mechanics of systems in thermal equilibrium has been a well-established part of the physics core curriculum. Yet, for natural systems, thermal equilibrium is the exception rather than the rule.  If we look around us, we see a world overwhelmingly far from equilibrium: from the intricate dynamics of living cells to more complex biological organisms; from ripples on a pond to global weather patterns.

The study of systems far from thermal equilibrium is both challenging and rewarding.  The reward comes from the chance to better understand the vast collection of non-equilibrium real-life systems. The challenge is obvious: we step outside the comfort of the familiar Gibbs \cite{Gibbs} framework for equilibrium systems, so we have to search for new tools to achieve fundamental understanding and theoretical classification of non-equilibrium behavior. Over the last three decades, an increasing number of condensed matter theorists are devoting their efforts to understanding complex collective behavior of far-from-equilibrium systems using methods that range from easily accessible computer simulations to sophisticated field theoretic techniques.

Non-equilibrium statistical physics is a relatively new field, not yet part of the physics curriculum at the undergraduate level.  Nonetheless, the methods and results of non-equilibrium statistical physics are being employed in fields as diverse as molecular biology, computer science, economics and politics.  Thus it seems suitable for undergraduate students of quantitative science to be introduced to the field early in their academic journey. 

This paper presents an introduction to the field of non-equilibrium statistical physics via a theoretical model which, while retaining the essence of the difficulties of far-from-equilibrium systems, is as simple as possible. The "two temperature kinetic Ising model" has two appealing features:   multi-temperature systems are fairly common, for instance a water tank with an immersion heater, nuclear magnetic resonance in an external magnetic field, or a lattice of nuclei in a solid prepared at a certain spin temperature \cite{Schmuser}; as well, the model can be solved analytically for small system sizes using only basic linear algebra, and can also be modeled via Monte Carlo simulations without too much difficulty. For these reasons, this model has great pedagogical value.  Undergraduate students in a traditional course in statistical physics can extend principles learned there to gain first exposure to far-from-equilibrium systems.

This paper is structured as follows:  We first (Section \ref{sect2}) give an overview of non-equilibrium statistical physics and compare it with its equilibrium counterpart.  Next, (Section \ref{sect3}) we introduce the "two temperature kinetic Ising model" as a simple example of a non-equilibrium system. Using the microscopic approach, we solve exactly the master equation for some small system sizes, to obtain the probability distribution (Section \ref{sect4}). We also comment on the probability currents (Section \ref{sect5}), a fundamental feature of far from equilibrium systems in the steady state. By direct calculation, we also find the time dependence of the probability distribution for a small system (Section \ref{sect6}). 

\section{In and out of equilibrium: comparison and contrast}\label{sect2}
Statistical mechanics is the discipline which bridges the gap between the microscopic world of molecules, atoms and electrons and the macroscopic world of thermodynamics and bulk properties of materials.  It enables us to predict macroscopic behavior of a system, knowing the quantitative properties of molecular interactions. The foundation of equilibrium statistical mechanics rests upon Boltzmann's fundamental hypothesis \cite{Boltzmann}. 
\begin{quote} If an isolated macroscopic system is ergodic \cite{Pathria}, it will reach thermal equilibrium after a sufficiently long time. Then every configuration (or microstate) available to the system can be found with equal probability. \end{quote}
Of course, completely isolated systems are rather unrealistic, so typically we consider systems in contact with a very large (infinite) heat reservoir. After a sufficiently long time equilibrium is reached, meaning that the average net energy flux between the system and the thermal bath vanishes and both have reached the same temperature, $T$. Under these conditions, the probability for finding the system in a microstate $\sigma $ is given by the canonical distribution:
\begin{equation}
P_{eq}(\sigma )=\frac{e^{-\beta H(\sigma )}}{Z}.  \label{Peq}
\end{equation}
where $\beta =1/(k_{B}T)$ is related to the inverse temperature via Boltzmann's constant $k_{B}$, and $Z$ is the factor that ensures the normalization of probabilities, known as the partition function.  Thus, once we have specified
a labeling of the microscopic configurations, $\{\sigma \}$, and we have determined the microscopic Hamiltonian $H(\sigma )$ --- i.e. the internal energy of each configuration --- so that the Boltzmann factor, $e^{-\beta H}$, is known, we can calculate, at least in principle, the partition function of the equilibrium system as well as average values of time-independent observables.  Of course, we can run into technical difficulties - in particular, some of the configurational sums may not be obtainable exactly, etc. - but the fundamental framework for a solution is fully established.

The jump from the idealization of thermal equilibrium to the full, possibly turbulent, dynamics of the real world is too great for our limited
present knowledge of complex non-equilibrium systems, and we are therefore well advised to focus on the simplest extension of equilibrium systems, namely, \emph{non-equilibrium steady states}. This category of systems is characterized by time-independent macroscopic behavior which results from applying a uniform driving force. We may model the external driver as a second temperature bath, at a different temperature than the first, thus feeding energy into the system or removing it. The long-time behavior of such a two-bath system exhibits constant energy flow through the system.  A simple example of such a system is a curent-carrying electrical resistor in a steady state, gaining energy from a current source and losing it as heat to the environment.  The resistor is not in equilibrium because there is a non-zero energy flux flowing through it; yet, after a sufficiently long time, it reaches a ``steady state'', with time-independent macroscopics. Understanding such a system in the steady state involves finding the associated stationary probability distribution of microstates, which is the long-time limit of the time-dependent microstate probability distribution.

A starting point for the study of these systems is the master equation, expressing the conservation of configurational probabilities. We consider a continuous-time dynamics, with a finite and discrete configuration space ${\sigma}$. The time-dependent probability $P(\sigma ,t)$ for finding the system in configuration $\sigma $ at time $t$ changes only due to transfer of probability into $\sigma $ from other configurations, or from $\sigma$ into others,  in such a way that $\sum_{\sigma }P(\sigma ,t)=1$ at all times.  The evolution of probability $P(\sigma ,t)$  is dictated by a set of transition rates $c\left[ \sigma \rightarrow \sigma ^{^{\prime }}\right]$ that describe the evolution of the system from one configuration $\sigma$ to a different configuration $\sigma^{^{\prime }} $, per unit time. For example, given a spin system, one configuration leads into another via a spin flip. We may write a balance (continuity) equation: its right hand side consists of two sums: the first is a ``gain'' term, summing over all configurations from which configuration $\sigma $ could possibly result while the second is a ``loss'' sum, accounting for all configurations into which $\sigma $ can evolve:
\begin{equation} 
\dfrac{dP(\sigma ,t)}{dt}=\sum_{\sigma ^{\prime }}\left\{ c\left[ \sigma
^{\prime }\rightarrow \sigma \right] P(\sigma ^{\prime },t)-c\left[ \sigma
\rightarrow \sigma ^{\prime }\right] P(\sigma ,t)\right\}   \label{master}
\end{equation}
Here, $c\left[ \sigma \rightarrow \sigma ^{^{\prime }}\right] $ denotes the transition rate from configuration $\sigma $ into another
configuration $\sigma ^{^{\prime }}$. These rates must be given as part of defining a specific model. Our task in solving the steady-state problem is to find the stationary solution of this equation, $P^{*}(\sigma )\equiv \lim_{t\rightarrow \infty }P(\sigma
,t)$, for which the left hand side of Eqn. (\ref{master}) vanishes:
\begin{equation}
\begin{array}{l}
0=\dfrac{dP^{*}(\sigma )}{dt}=\sum_{\sigma ^{^{\prime }}}\left\{ c\left[
\sigma ^{^{\prime }}\rightarrow \sigma \right] P^{*}(\sigma ^{^{\prime
}})-c\left[ \sigma \rightarrow \sigma ^{^{\prime }}\right] P^{*}(\sigma
)\right\} 
\end{array}
\end{equation}
Of course, the steady state distribution $P^{*}(\sigma )$ will depend on the transition rates. Under rather general conditions on the $c$'s, this solution will be unique, and thus independent of initial conditions.

For a system in thermal equilibrium with a single heat bath, we know its steady state distribution must be given by $P_{eq}(\sigma )$ from Eq. (\ref{Peq}).  For the equilibrium case, we must thus choose the rate terms $c\left[ \sigma \rightarrow \sigma ^{^{\prime }}\right] $ to be consistent with the canonical distribution result. This requirement forces the rates to satisfy the "detailed balance" condition, 
\begin{equation}
\dfrac{c\left[ \sigma ^{^{\prime }}\rightarrow \sigma \right] }{c\left[
\sigma \rightarrow \sigma ^{^{\prime }}\right] }=\dfrac{P_{eq}(\sigma )}{%
P_{eq}(\sigma ^{^{\prime }})}  \label{db}
\end{equation}

Since $ P_{eq}(\sigma )\propto \exp (-\beta H)$, we therefore choose our rates such that
\begin{equation}
\begin{array}{l}
\dfrac{c\left[ \sigma ^{^{\prime }}\rightarrow \sigma \right] }{c\left[
\sigma \rightarrow \sigma ^{^{\prime }}\right] }=\exp (\beta \Delta H)
\end{array}
\end{equation}
where
\begin{equation}
\begin{array}{l}
\Delta H=H(\sigma ^{^{\prime }})-H(\sigma )
\end{array}
\end{equation}

For systems in equilibrium, the "detailed balance" condition is an intrinsic property, and is related to their microscopic reversibility \cite{zia(paper)}. This condition assures the invariance of the long time limit of the probability distribution. This constraint on the transition rates is necessary when modeling (via Monte Carlo simulations, for instance) systems in thermal equilibrium.  A key feature of a system far from thermal equilibrium is the violation of detailed balance: its steady state distribution $P^{*}(\sigma )$ does not
satisfy Eqn.(\ref{db}).  For a non-equilibrium system in its steady state, the question becomes how one generalizes the detailed balance condition such that, when the "drive" is turned off, the equilibrium solution is recovered.

A more intuitive way to discuss the detailed balance condition is to describe it in terms of probability currents \cite{zia(paper)}. We consider a series of configurations $\sigma _{1},\sigma _{2},..\sigma _{n}$,
which form a cycle, each successive state reachable in a single step from its predecessor (likewise for states $\sigma_n$ and $\sigma_1$):  We define the products of the rates around the cycle as:
\begin{eqnarray}
\Pi _{+} &=&c\left[ \sigma _{1}\rightarrow \sigma _{2}\right] c\left[
\sigma _{2}\rightarrow \sigma _{3}\right] \cdots c\left[ \sigma _{n}\rightarrow
\sigma _{1}\right]  \\
\Pi _{-} &=&c\left[ \sigma _{2}\rightarrow \sigma _{1}\right] c\left[
\sigma _{3}\rightarrow \sigma _{2}\right] \cdots c\left[ \sigma _{1}\rightarrow
\sigma _{n}\right]   \nonumber
\end{eqnarray}

Detailed balance holds if and only if 
\begin{equation}
\Pi _{+}=\Pi _{-}
\end{equation}
for \emph{all }cycles, which is equivalent to saying that the net probability current between any two configurations vanishes in the steady state:
\begin{equation}
c\left[ \sigma ^{^{\prime }}\rightarrow \sigma \right] P^{*}(\sigma
^{^{\prime }})-c\left[ \sigma \rightarrow \sigma ^{^{\prime }}\right]
P^{*}(\sigma )=0
\end{equation}

If the rates violate the detailed balance condition, then there will be non-trivial current loops:
\begin{equation}
c\left[ \sigma ^{^{\prime }}\rightarrow \sigma \right] P^{*}(\sigma
^{^{\prime }})-c\left[ \sigma \rightarrow \sigma ^{^{\prime }}\right]
P^{*}(\sigma )\neq 0
\end{equation}
The presence of these current loops is a key characteristic of non-equilibrium steady states, and a signal of the microscopic irreversibility of these systems. For a complete and unique characterization of a non-equilibrium steady-state one needs to specify both the configurational probability distribution and the distribution of the probability currents. \cite {zia(paper)} The choice of the transition rates is not unique. The transition rates we will use (see \ref{sect3}) for our two-temperature kinetic model have been previously studied \cite {Glauber} and are known to lead to the Ising model solution for the equilibrium case. We will examine the specific probability currents for our model in order to illustrate the profound differences between equilibrium and non-equilibrium systems.\rm  

From a pedagogical point of view, these calculations should provide students with a deeper understanding of the profound difference between equilibrium and non-equilibrium statistical physics. Starting "from scratch", at the microscopic level, students have the opportunity  to learn important statistical physics concepts such as: choice of transition rates, balance equations, steady state, probability distributions, equivalence classes and boundary conditions. Students who are familiar with equilibrium statistical physics methods are now faced with the challenge of being outside of the traditional framework of the canonical distribution, in search of new methods of study. Also, these analytical calculations complement straightforward computer simulations of nonequilibrium states.  

In the following section, we introduce our model and present some sample calculations and results for an $N = 4$ lattice. It is quite instructive to see the transformation in configurational probabilities as the non-equilibrium condition is "turned on" by gradually allowing two reservoirs that permit energy flow in the system to take on different temperatures.

\section{The two temperature kinetic model}\label{sect3}

The Ising model was introduced by Lenz in 1925 \cite{Lenz, Ising}, in an attempt to understand the nature of phase transitions in ferromagnets. It has become a paradigm of statistical physics, and is a common feature of statistical physics classes at the undergraduate level. Building a non-equilibrium model related to the Ising model is a good pedagogical approach -- students already familiar with the Ising model can quickly come to appreciate the novel behaviors of systems that are far from equilibrium.  Paralleling the standard definition of the Ising model, we define our one-dimensional system as follows.  Consider a collection of adjacent sites numbered $i=1,2,\ldots,N$, with site $N$ considered adjacent to site $1$, as if the points were distributed around a ring (so-called "periodic boundary condition".)  Each site $\ i$ can be full or empty, and in order to describe a particular system configuration we define a set of occupation numbers $n_{i}$, each being $0$ for an empty site and $1$ for an occupied site.  We will alternately use spin language notation $\sigma _{i}=2n_{i}-1$, with $\sigma_{i}=1$ for a full cell and $\sigma _{i}=-1$ for an empty cell. We assume an even number of sites; periodic boundary conditions imply that $\sigma _{N+1}=\sigma _{1}.$ The spins are in contact with \emph{two} heat baths at \emph{different} temperatures $T_{e}$ and $T_{o}$, in such a way that spins on even lattice sites experience $T_{e}$ and those on odd sites, $T_{o}$. Imposing $T_{e}\neq T_{o}$ drives the system out of equilibrium: each heat bath tries to drive the system towards equilibrium with the same Hamiltonian but at its own temperature. As a result, energy flows from one sublattice to the other and the steady state is a nonequilibrium one.

We endow the spins with nearest-neighbor interactions, according to the usual Ising Hamiltonian:
\begin{equation}
\begin{tabular}{l}
$H=-J\sum_{i}\sigma _{i}\sigma _{i+1}$,
\end{tabular}
\end{equation}
where $J$ represents half the energy difference between a state of two adjacent parallel spins and one of two adjacent opposite spins.  The dynamics is modeled by a generalization of the Glauber rates \cite{Glauber}. The $n$-th spin is flipped on a time scale of $\tau$ with a rate
\begin{equation}
c_{n}\left( \left\{ \sigma \right\} \right) =\dfrac{1}{2\tau }\left( 1-%
\dfrac{\gamma _{n}}{2}\sigma _{n}\left( \sigma _{n+1}+\sigma _{n-1}\right)
\right) 
\label{w-2T}
\end{equation}
where $0\leq \gamma _{n}\leq 1$ is related to the local temperature
according to

\begin{equation}
\gamma _{n}=\left\{ 
\begin{array}{c}
\tanh (\dfrac{2J}{k_{B}T_{e}}) \\ 
\\ 
\tanh (\dfrac{2J}{k_{B}T_{o}})
\end{array}
\right. \quad \text{for\quad }
\begin{array}{c}
n\text{ even} \\ 
\\ 
n\text{ odd}
\end{array}
\end{equation}

These rates are invariant under a global spin flip $\{\sigma \}\rightarrow \{-\sigma \}$ and under translations by an even number of sites $\sigma _{n}\rightarrow \sigma _{n+2j}$ for all $n$, $j=1,2,..N$  (translational invariance modulo 2). Therefore, the same invariance should hold for the steady state distribution $P^{*}\left( \left\{ \sigma \right\} \right) $ , namely, $P^{*}\left( \left\{ \sigma \right\} \right) =P^{*}\left( \left\{
-\sigma \right\} \right) $ and $P^{*}\left( \left\{ \sigma _{n}\right\}
\right) =P^{*}\left( \left\{ \sigma _{n+2j}\right\} \right) $  This will allow us to reduce the number of configurations considered in our model.

When the two heat baths have the same temperature, $T_{e}$ $=T_{o}=T$, the above Hamiltonian and rates define the exactly solvable Glauber model \cite{Glauber} which relaxes to the equilibrium state of the Ising model.

The non-equilibrium version of this one-dimensional kinetic model has been previously studied \cite{RZ(kinetic),Schmuser,Mobilia}. The magnetization (average over all spins) and the two-spin correlations (average over all pairs of spins) were calculated for both the steady state and the time-dependent case.\cite{RZ(kinetic),Schmuser, Mobilia}. So far, there is no compact expression for the steady-state probability distribution.  Here, we exhibit an exact expression for the \emph{full }probability distribution, but at the price of restricting ourselves to very small systems. 

Below we show the principal steps of the calculations and the most significant results. 

\section{Probability distribution}\label{sect4}

The master equation tells us how a particular configuration evolves in time:

\begin{equation}
\partial_t P\left( \{\sigma \},t\right) 
=\sum_{n=1}^{N}\left[ c_{n}(\{\sigma ^{[n]}\})P\left( \{\sigma
^{[n]}\},t\right) -c_{n}(\{\sigma \})P\left( \{\sigma \},t\right) \right] 
\label{meq}
\end{equation}
where the state $\{\sigma ^{[n]}\}$ differs from $\{\sigma \}$ by a flipping
of the $n$-th spin and the rates are given by Eqn. (\ref{w-2T}). We seek the
steady state solution:

\begin{equation}
\begin{tabular}{l}
$P^{\ast }\left( \left\{ \sigma \right\} \right) \equiv \lim_{t\rightarrow
\infty }P\left( \{\sigma \},t\right) $%
\end{tabular}
\end{equation}

To simplify the problem, we note that configurations which are related by symmetries of the dynamics will obviously occur with the same probability. For example, the configurations $+-++$ and $+++-$ will have the same probability in an $N = 4$ system, since they result from each other by a translation modulo 2. Similarly, $+-++$ and $-+--$ are related by a
global spin flip. We can therefore define ``equivalence classes'', each class consisting of those configurations related to one another by a symmetry transformation. We need only solve for the probability associated with one member of each equivalence class, thus greatly reducing the number of unknowns below the total number of configurations, $2^{N}$. 

Below, we pursue this program for an $N = 4$ system, present the calculations in some detail. We also show the results for an $1N = 6$ system. The exact stationary probabilities will be determined and studied as a function of a\ single
parameter, $\gamma _{e}$ for some fixed, suitably chosen value of $\gamma _{o}$. Clearly, the equilibrium limit is represented by $\gamma _{e}=\gamma _{o}$.

\begin{figure}[p]\label{equivclass}
  \centering
  \includegraphics[bb=15 13 597 780,width=5in,height=6.59in,keepaspectratio]{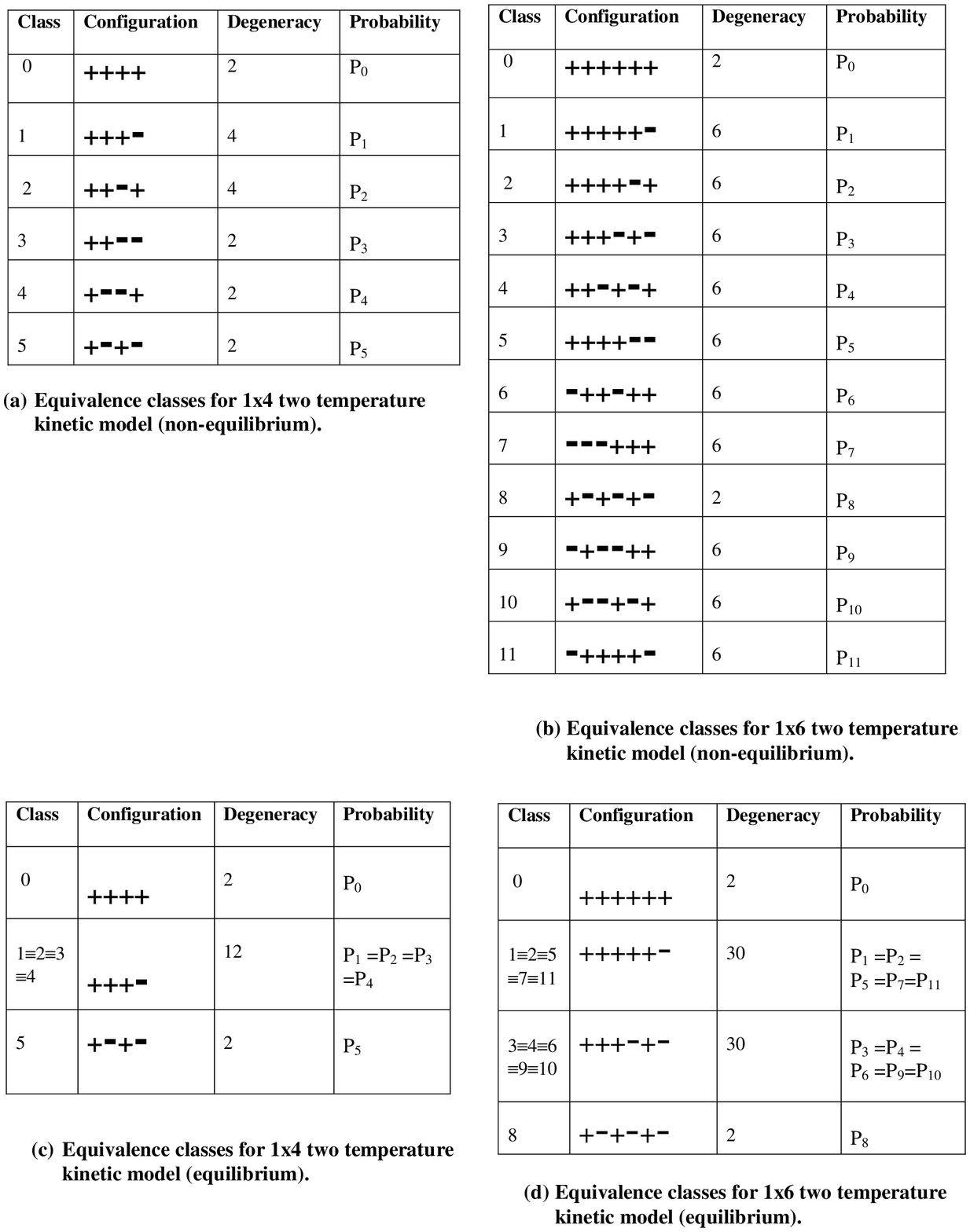}
  \caption{Equivalence classes for 1x4 and 1x6 systems}
  \label{fig:classes_new}
\end{figure}

For an $N = 4$ system, there are 6 equivalence classes, numbered (in some arbitrary order) $i=0,1,...,5$. The degeneracy $d_{i}$ of each class is defined as the number of configurations in this class. $P_{i}$ denotes the
stationary probability associated with class $i$. In Fig. \ref{equivclass} we present the equivalence classes with one representative of each class. For comparison, the 12 equivalence classes for the $N = 6$ system are also listed.

To write the steady state equation for $P_{0}$, we need to identify the ``neighbors'' of class $0$ in configuration space, i.e., those configurations which can be reached from $P_{0}$ via a single spin flip, and vice versa: these configurations obviously belong to classes $1$ or $2$. Similarly, class $1$ is a neighbor to classes $0$, $3$, $4$ and $5$.  The master equation leads us to a set of six rate equations:
\begin{eqnarray}
  2\tau \partial _{t}P_{0} &=& 2(1+\gamma _{e})P_{1}+2\left( 1+\gamma
_{o}\right) P_{2}-2 \left( 2-\gamma _{e}-\gamma _{o}\right) P_{0} \nonumber \\
2\tau \partial _{t}P_{1} &=&(1-\gamma _{e})P_{0}+ P_{3}+P_{4}+\left( 1+\gamma _{e}\right) P_{5}-4P_{1} \nonumber \\
2\tau \partial _{t}P_{2} &=&(1-\gamma _{o})P_{0}+ P_{3}+P_{4}+\left( 1+\gamma _{o}\right) P_{5}-4P_{2} \nonumber \\
2\tau \partial _{t}P_{3} &=& 2P_{1}+ 2P_{2}-4P_{3} \nonumber \\
2\tau \partial _{t}P_{4} &=& 2P_{1}+ 2P_{2}-4P_{4} \nonumber \\
2\tau \partial _{t}P_{5} &=& (1-\gamma _{o})P_{2}+\left( 1-\gamma_{e}\right) P_{1}-
\left(2+\gamma _{o}+\gamma _{e}\right) P_{5} . \label{1x4master}
\end{eqnarray}
Since our probabilities are normalized, we have one additional equation, namely
\begin{equation}
1=\sum_{i=0}^{5}d_{i}P_{i} \label{norm}
\end{equation}

The probabilities for the equilibrium case $\gamma _{o}=\gamma _{e}\equiv \gamma $ are given by the Boltzmann factor, $\exp (-\beta H)$. In this case, probabilities differ only if their configurational energies are distinct, producing an even greater reduction in the number of distinct cases to consider. With a little algebra, one can convert the exponentials into functions of $\gamma $ to arrive at:
\begin{eqnarray}
P_{1}=P_{2}=P_{3}=P_{4}= P_{0}\dfrac{(1-\gamma )}{\left( 1+\gamma \right) } \nonumber \\ 
P_{5} = P_{0}\dfrac{(1-\gamma )^{2}}{\left( 1+\gamma \right)
^{2}} 
\end{eqnarray}

Using the normalization condition (Eq. \ref{norm}), we find for $P_{0}$ the following expression:
\begin{equation}
P_{0}=\dfrac{1}{8}\dfrac{(1+\gamma )^{2}}{\left( 2-\gamma ^{2}\right) }
\end{equation}

These equilibrium results can also be found by solving the rate equations, Eq. \ref{master}, with the left-hand sides set to zero, along with Eq. \ref{norm}.

\begin{figure}[p]
  \centering
  \includegraphics[bb=15 13 597 780,width=5.72in,height=6.5in,keepaspectratio]{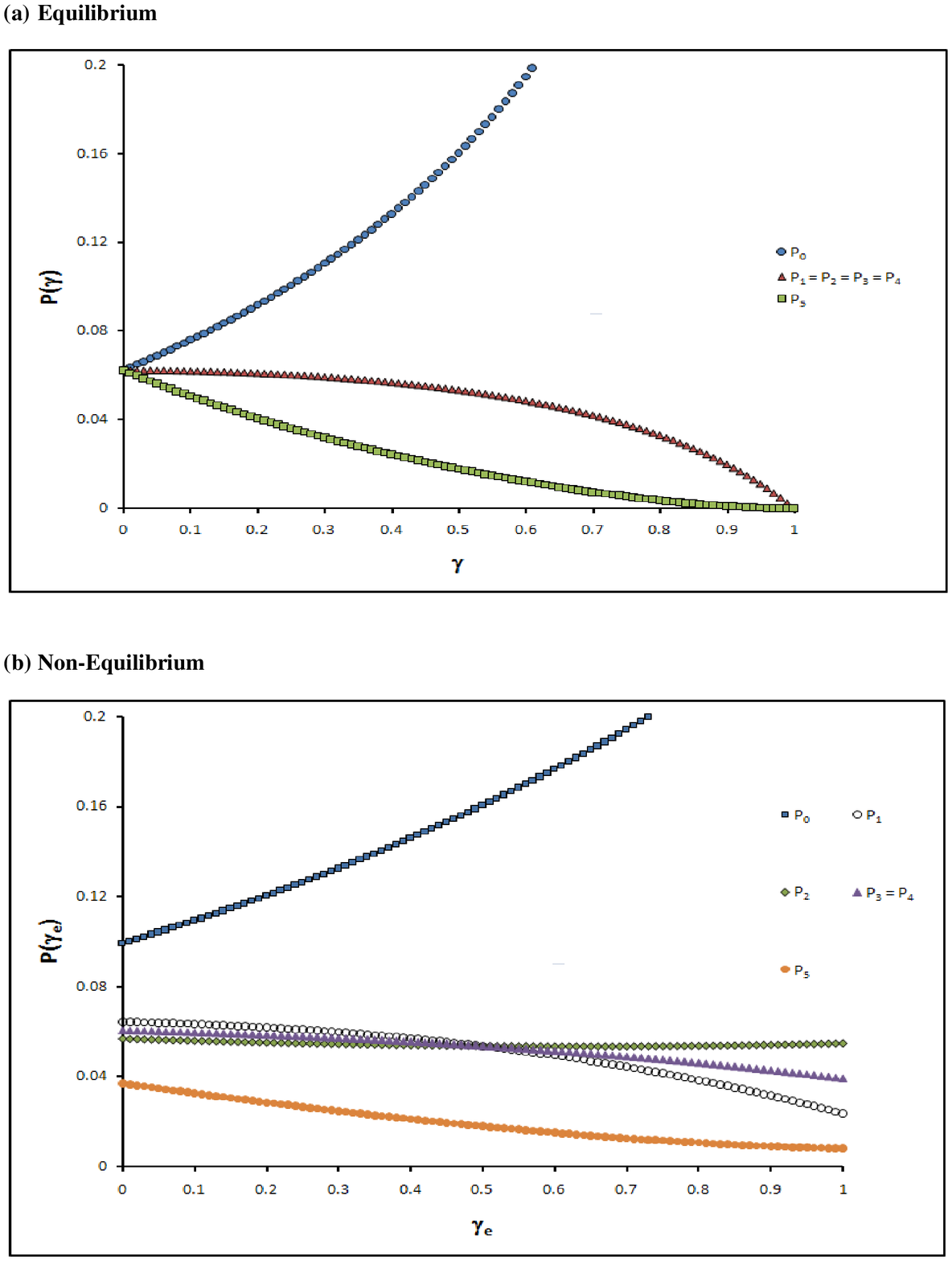}
  \caption{(a) Probability distribution of the 1x4 system as a function of $\gamma=\tanh (\dfrac{2J}{k_{B}T})$ in the equilibrium case, when even and odd sites are in contact with heat baths at the same temperature; (b) The probability distribution of the 1x4 system as a function of $\gamma_{e}$ in the driven case, when even and odd sites are in contact with heat baths at different temperatures, with $\gamma_{o}=0.5$.}
  \label{fig:prob1x4_11_15}
\end{figure}

\begin{figure}[p]
  \centering
  \includegraphics[bb=15 13 597 780,width=5.72in,height=6.5in,keepaspectratio]{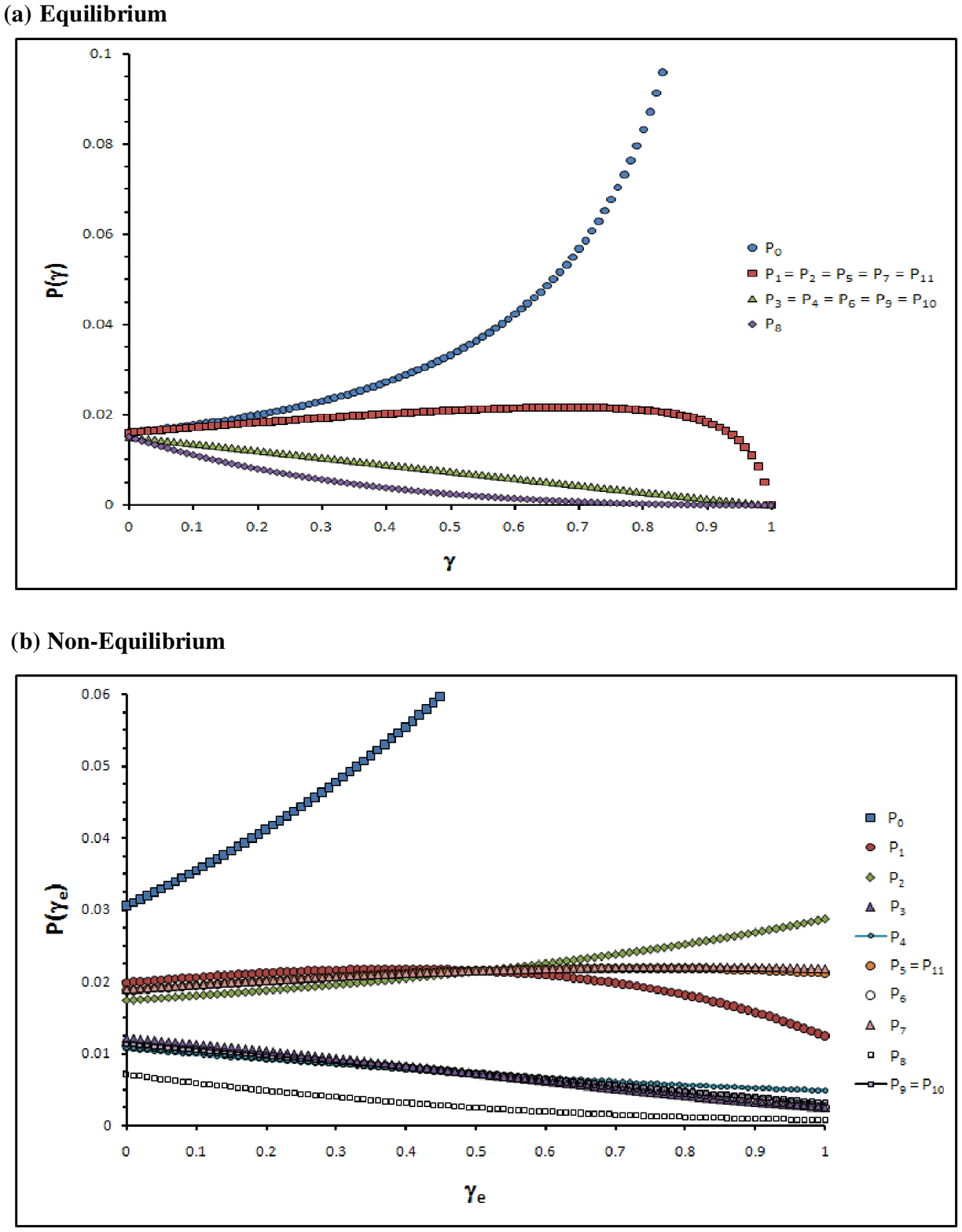}
  \caption{(a) Probability distribution of the 1x6 system as a function of $\gamma=\tanh (\dfrac{2J}{k_{B}T})$ in the equilibrium case, when even and odd sites are in contact with heat baths at the same temperature; (b) The probability distribution of the 1x6 system as a function of $m=\dfrac{\gamma_{e}}{\gamma_{o}}$ in the driven case, when even and odd sites are in contact with heat baths at different temperatures, with $\gamma_{o}=0.5$.}
  \label{fig:prob1x6_11_15}
\end{figure}

As we can see, at equilibrium only three different probabilities remain,
reflecting the three possible values of configurational energy: (i) no broken
(i.e., $+-$) bonds -- class $0$; (ii) two broken bonds -- classes $1, 2, 3, 4$
(iii) four broken bonds -- class 5. In Fig. \ref{fig:prob1x4_11_15}.a we show their dependence on $\gamma $. All
probabilities become equal for $\gamma \rightarrow 0$ which corresponds to
the infinite temperature limit, and all except $P_{0}$ vanish for $\gamma \rightarrow 1$, i.e., $T\rightarrow 0$. 
The situation for the $N = 6$ system, shown in Fig. \ref{fig:prob1x6_11_15}.a  is similar.  There are four distinct probabilities, all of which are zero at infinite temperature, and three of which vanish at zero temperature.

The problem becomes more difficult when equilibrium is not assumed, i.e., $\gamma _{o}\neq \gamma _{e}$.  It becomes necessary to solve simultaneously for the probabilities.  Steady state solutions are achieved by considering the equations, Eq. \ref{master}, resulting from the master equation, with the time partials all set to zero.  The normalization condition, Eq. \ref{norm}, implies that the six equations of Eq. \ref{master} must be redundant.  A solution emerges if one solves the inhomogeneous set of linear equations resulting from taking five of the equations from 
Eq. \ref{master} along with Eq. \ref{norm}.  This is easily done using algebraic software, and
 produces the results
\begin{eqnarray}
   P_1 &=& \frac{ 8+\gamma_o^2-6\gamma_o  \gamma_e -3 \gamma_e^2}{64(2-\gamma_o \gamma_e)} \nonumber \\
   P_2 &=& \frac{8 - 3 \gamma_o^2 - 6 \gamma_o \gamma_e + \gamma_e^2}{64(2-\gamma_o \gamma_e)}
\nonumber \\
   P_3 &=& \frac{8- \gamma_o^2 - 6 \gamma_o \gamma_e - \gamma_e^2}{64(2-\gamma_o \gamma_e)}
\nonumber \\
   P_4 &=& \frac{8- \gamma_o^2-6 \gamma_o \gamma_e - \gamma_e^2}{64(2-\gamma_o \gamma_e)}
\nonumber \\
   P_5 &=& \frac{8-3 \gamma_o^2 -2 \gamma_o \gamma_e -3 \gamma_e^2+8 \gamma_o +8 \gamma_e}{64(2-\gamma_o \gamma_e)}
\nonumber \\
   P_0 &=& \frac{8+3 \gamma_o^2 +2 \gamma_o \gamma_e +3 \gamma_e^2+8 \gamma_o +8 \gamma_e}{64(2-\gamma_o \gamma_e)} .
\label{1x4ss}
\end{eqnarray}
We exhibit these probabilities in Fig. \ref{fig:prob1x4_11_15}.b, where we plot each probability vs $\gamma_{e}$, for $\gamma _{o}=0.5$, letting $\gamma_{e}$ run through its full range, $0$ to $1$.  First of all, there are more distinct probabilities than for the equilibrium case. The equations show that each equivalence class behaves differently, except for the added degeneracy that results from the dynamics we have chosen: $P_3=P_4$.  The crossings seen at the point where $\gamma_e = \gamma_0$ simply reproduce the equilibrium results for $\gamma = .5$.  We also note that $P_{0}$ remains the
most probable configuration. Finally, we observe a grouping of curves: The probabilities of configurations which
share equal configurational energy track each other quite closely, and those associated with \emph{different} configurational energy\emph{\ never cross}. One might be tempted to conjecture that this is a generic feature of this
simple non-equilibrium system. Unfortunately, it does not persist for larger system sizes, such as $N = 8$. This study also allows us to seek configurations, unrelated by symmetry, which would nevertheless occur with the same probability. In equilibrium, such configurations would all have the same energy, i.e., degeneracies (beyond symmetry) are controlled by energy. Far from equilibrium, it is not known which quantity controls such degeneracies. The difference between equilibrium and non-equilibrium probabilities is quite dramatic, and any connections between the two are far from obvious.

\section{Probability currents}\label{sect5}

The presence of current loops marks a fundamental difference between non-equilibrium and equilibrium systems, and we here illustrate this for our $N = 4$ system. In Section \ref{sect2} we defined these probability current loops as:

\begin{equation}
   J\left[\sigma ^{\prime}\rightarrow \sigma \right]  =  c\left[ \sigma ^{^{\prime }}\rightarrow \sigma \right] P^{*}(\sigma
^{^{\prime }})-c\left[ \sigma \rightarrow \sigma ^{^{\prime }}\right]
P^{*}(\sigma )\neq 0
\end{equation}

\begin{figure}[p] 
  \centering
  \includegraphics[bb=15 13 597 780,width=4.3 in,height=6.00in,keepaspectratio]{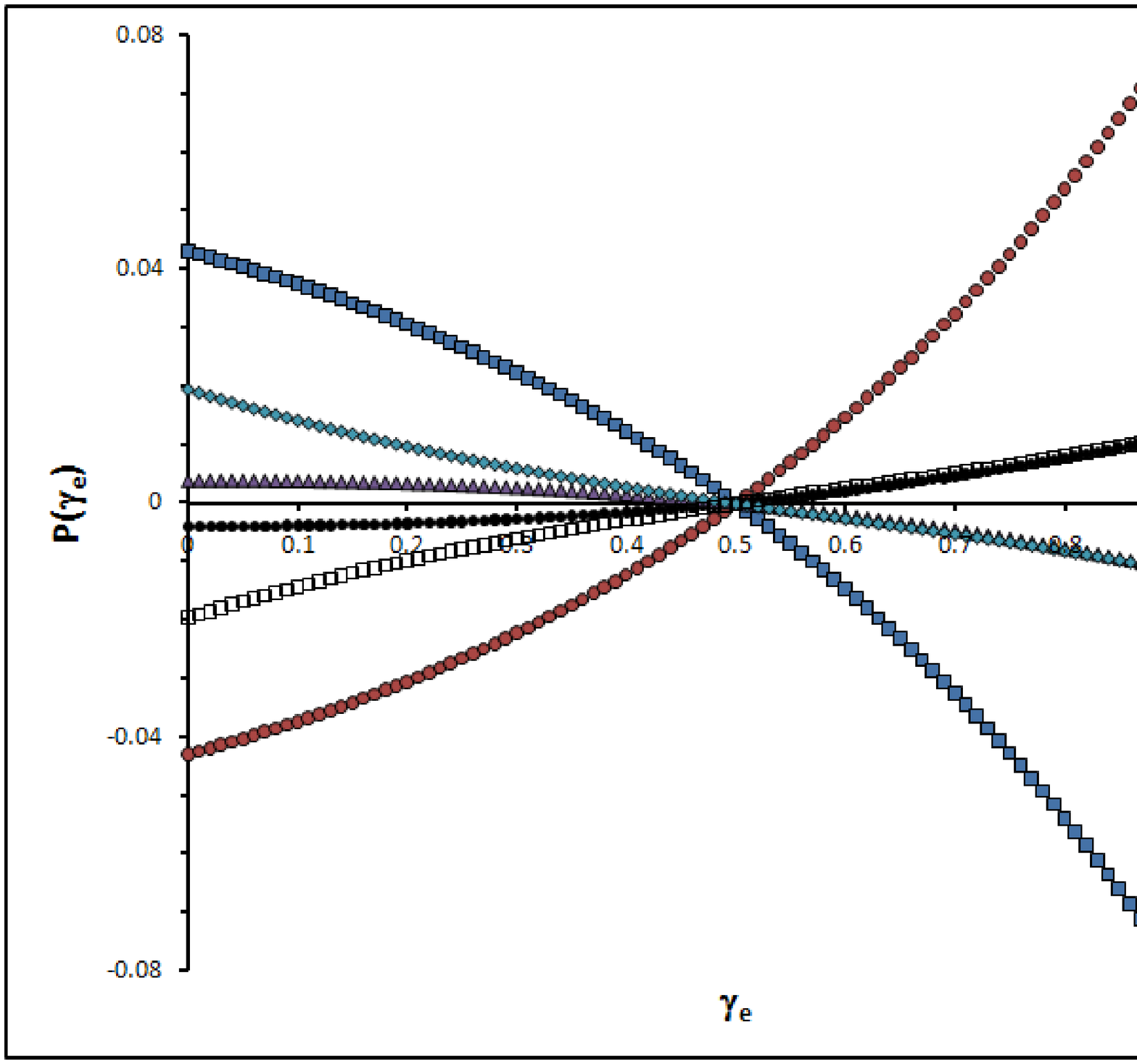}
  \caption{ The probability currents for the 1x4 system as a function of $\gamma_{e}$ in the driven case, when the even and odd sites are in contact with heat baths at different temperatures, with $\gamma_{o}=0.5$.}
  \label{probcur}
\end{figure}

In the equilibrium case, due to the detailed balance condition, these current loops vanish for any pair of configurations in the steady state. For our $N = 4$ system in steady state, we can calculate these currents as:
\begin{eqnarray}
J_{01}=(1-\gamma_{e})P_{0}-(1+\gamma_{e})P_{1}\\
J_{02}=(1-\gamma_{o})P_{0}-(1+\gamma_{o})P_{2}\\
J_{13}=J_{14}=P_{1}-P_{3}\\
J_{23}=J_{24}=P_{2}-P_{3}\\
J_{15}=(1-\gamma_{e})P_{1}-(1+\gamma_{e})P_{5}\\
J_{25}=(1-\gamma_{o})P_{2}-(1+\gamma_{o})P_{5}\\
\end{eqnarray}
 where $J_{ik}$ represents the current between configurations "i" and "j", where $i,j=0 \ldots 5$. Fig. \ref{probcur} shows these currents as a function of $\gamma_{e}$, for $\gamma_{o}=0.5$. We can see how all currents vanish at equilibrium, where $\gamma_{e}=\gamma_{o}=0.5$, in accordance with the detailed balance condition.

\section{Time Dependence}\label{sect6}
In principle, the present model can be fully and analytically solved for the time dependence of the configuration probabilities, given any initial set of probabilities as an initial condition.  The master equation leads to a set of coupled, first-order differential equations in time that can be solved using standard methods.  However the size of the system of equations grows very quickly with the number of cells, making such a straightforward approach difficult and rendering computer simulations a much more reasonable approach.  Nonetheless, it is instructive to look at one nontrivial case where complete solutions can be found:  the $N = 4$ case.  

As we have seen, this case allows $2^{4}=16$ distinct spin configurations, leading to $16$ distinct time-dependent probability functions.  When examining asymptotic time behavior the probabilities group into six equivalence classes, leaving a very tractable $6\times6$ linear algebraic system to solve for asymptotic probabilities.  For arbitrary times, this simplification is not available, since the initial conditions need not obey the symmetries of the asymptotic situation.  

It is a straightforward task to set up the $16$ time-dependent equations for the probabilities that follow from the master equation.  They can be simply represented by
\[ 2 \tau \partial_t{\vec{P}} = \mathcal{A} \vec{P}, \]
where $\vec{P}$ is a $16$ component vector whose components are the configurational probabilities, and $\mathcal{A}$ is the matrix of coefficients of the probabilities from the right-hand side of the master equation.  If we assume a time dependence for the vector $\vec{P}$ of $\exp{(-\lambda)}$ the equation reduces to the eigenvalue equation for the matrix $\mathcal{A}$:  its eigenvalues $a_i$  are proportional to the decay coefficients $\lambda$:  $\lambda_i = a_i /(2\tau)$.  One of the standard algebraic software packages, like Maple, can easily identify the eigenvalues for this system, as a function of the parameters $\gamma_e$ and $\gamma_o$, as $0,2,4,6,8,2(1\pm \sqrt{\gamma_e \gamma_o}),4(1\pm \sqrt{\gamma_e \gamma_o}), 6(1\pm \sqrt{\gamma_e \gamma_o})$. The eigenvalues $2$ and $6$ are both twofold degenerate, and $4$ is fourfold degenerate.  It is worthy of notice that these eigenvalues are coincident with the set given by Glauber for the one-temperature case of this model, in the limit $\gamma_e = \gamma_o$.

The general time-dependent solution is
\[  \vec{P}(t) = \sum_i c_i \vec{V}_i \exp{(-\lambda_i t)}, \]
where $\vec{P}(t)$ is a $16$ component vector carrying the time dependence of each configuration probability, $\vec{V}_i$ is the eigenvalue of $A$ corresponding to eigenvalue $\lambda_i$, and the sum runs over all eigenvectors.  In the case of degenerate eigenvalues, a linearly independent set of eigenvectors is chosen.  The constants $c_i$ are fully determined by the initial probabilities:
\[ \vec{C} = V^{-1} \vec{P}_i, \]
where $\vec{C}$ is the vector with components $c_i$, $V$ is a matrix whose columns are the eigenvectors of $A$, and $\vec{P}_i$ is the vector with components equal to the initial probabilities.

From the general solution it is clear that the asymptotic time behavior results solely from the term corresponding to the $\lambda = 0$ eigenvalue.  The associated eigenvector, normalized, coincides with the collection of probabilities derived earlier for the steady-state solution.  

\begin{figure}[p] 
  \centering
  \includegraphics[width=5.67in,height=4.97in,keepaspectratio]{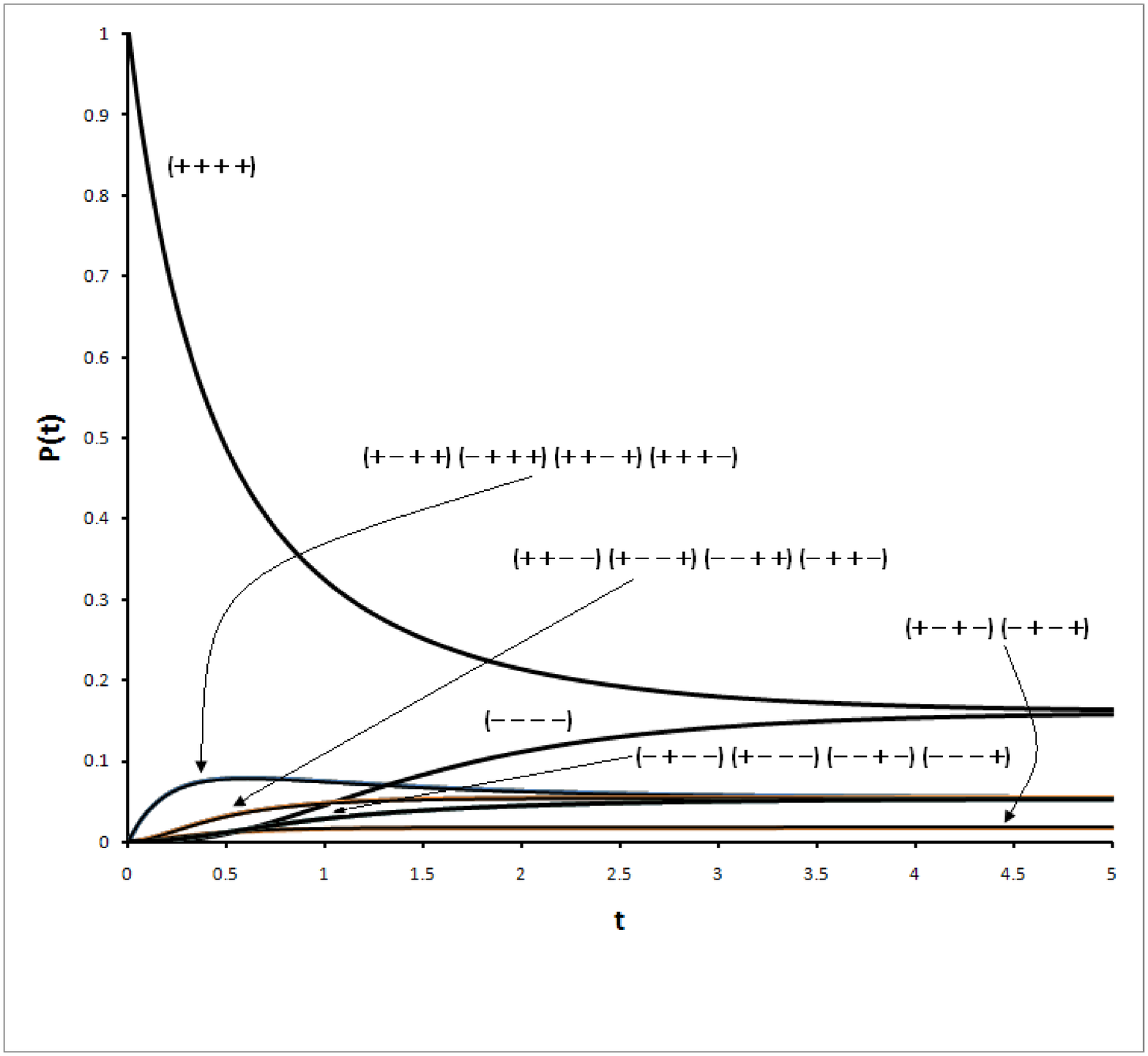}
  \caption{Time dependence of configuration probabilities in the $N = 4$ case for $\gamma_e = \gamma_o = .5$, with the ++++ configuration as the initial condition.  Vertical axis is absolute probability, time units are arbitrary.}
  \label{td1}
\end{figure}

\begin{figure}[p] 
  \centering
  \includegraphics[width=5.67in,height=5.38in,keepaspectratio]{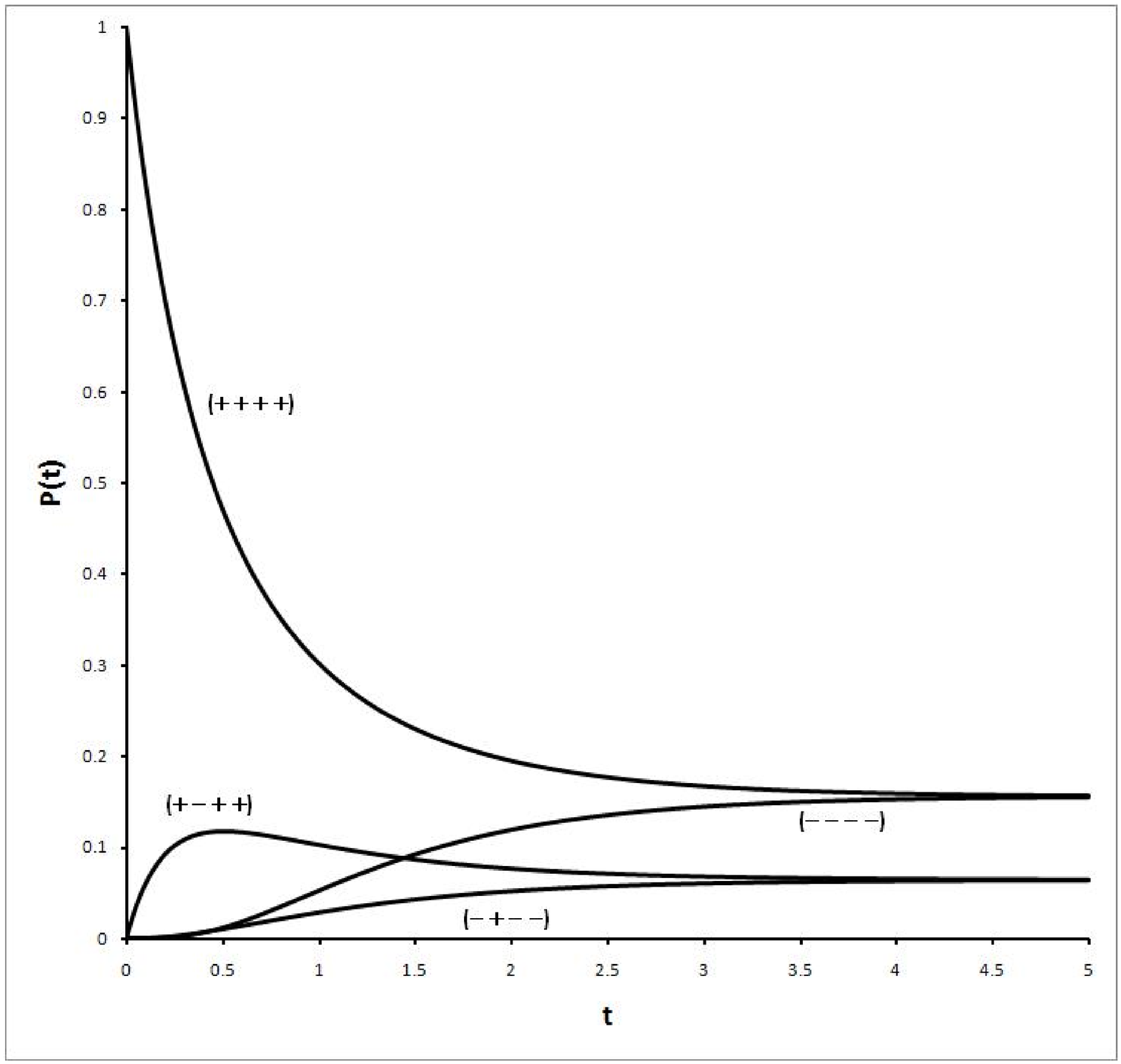}
  \caption{Time dependence of configuration probabilities in the $N = 4$ case for $\gamma_e = .2,  \gamma_o = .8$, with the ++++ configuration as the initial condition.  Vertical axis is absolute probability, time units are arbitrary.   (a)}
  \label{td2}
\end{figure}

\begin{figure}[] 
  \centering
  \includegraphics[width=5.67in,height=5.38in,keepaspectratio]{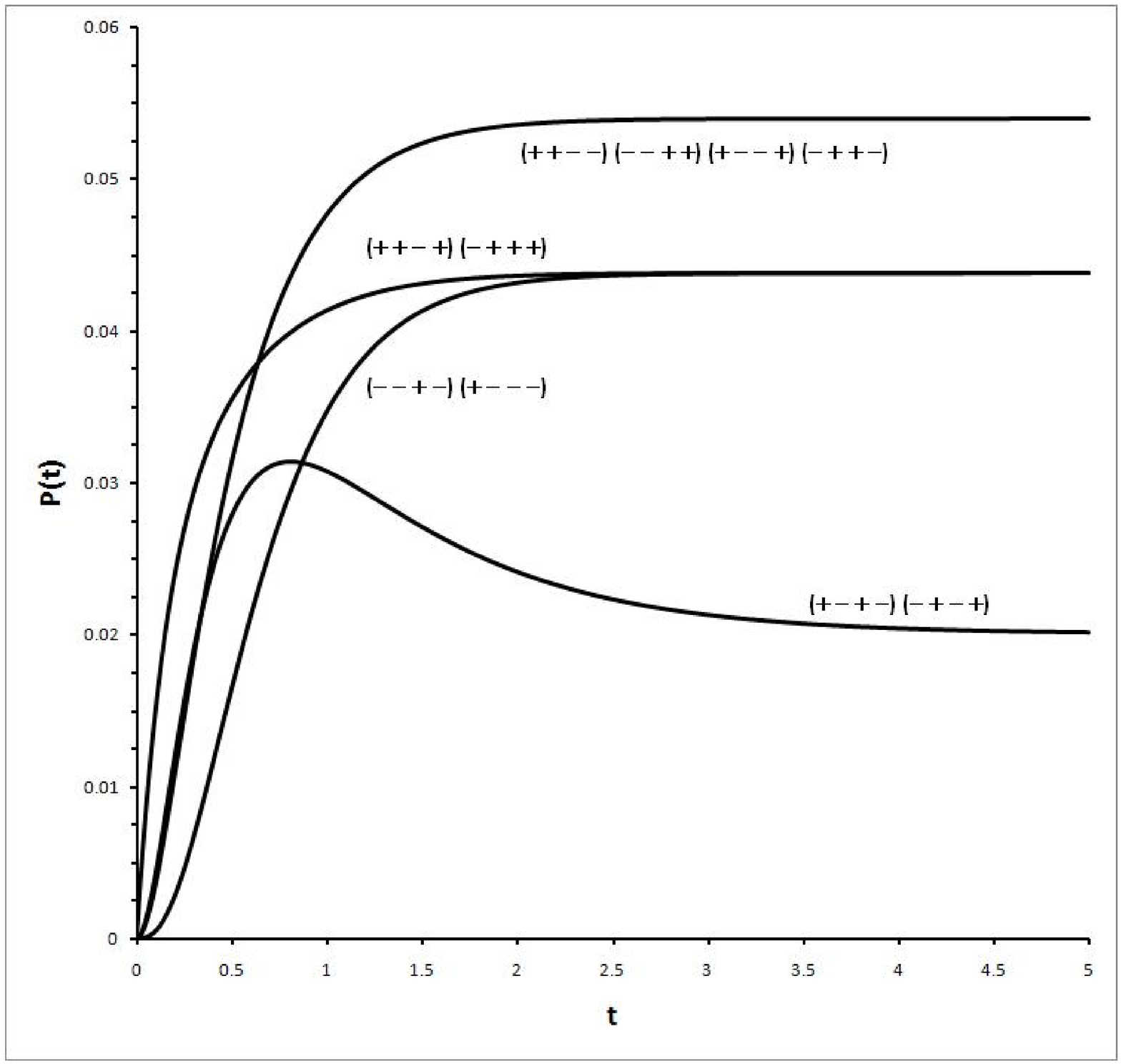}
  \caption{Time dependence of configuration probabilities in the $N = 4$ case for $\gamma_e = .2,  \gamma_o = .8$, with the ++++ configuration as the initial condition.  Vertical axis is absolute probability, time units are arbitrary.   (b)}
  \label{td3}
\end{figure}

By way of example, we present graphs of time-dependent behavior of the $N = 4$ system in two specific examples. In both cases, the system at $t=0$ is fully in the configuration ++++.   In the first, simpler case, we look at the situation in which the two temperature baths are at the same temperature, so that the eventual behavior of the system tends to equilibrium.  Time dependence of the probabilities for this system are exhibited in Fig. \ref{td1}.  Because of the simplicity of the initial state and the symmetry introduced by the equal temperatures, only five of the sixteen configuration probabilities show distinct time behavior.  At large values of t, these five probabilities collapse into three values corresponding to the equivalence classes exhibited in Fig. \ref{equivclass}.(c).  The fact that even in this simple case some probabilities grow from zero to a maximum before falling to their asymptotic values indicates some nontrivial physical effects are possible.  Looking at the same initial condition, but for two distinct temperatures ($\gamma_e = .2, \gamma_o = .8$), we are examining a system that never reaches equilibrium, but does eventually achieve a steady state, as discussed earlier. As shown in Fig.s \ref{td2} and \ref{td3}, eight of the sixteen configurations evolve distinctly, eventually reaching the steady state of the six equivalence classes listed in Fig. \ref{equivclass}.a.  As for the former example, there are configuration probabilities that grow or decay monotonically from their initial to final values, but others that initially grow past their steady state values before relaxing to them.  Since one can describe all physical properties of this system as functions of the configuration probabilities, this behavior suggests the possibility of time-dependent peaks in some of these properties.  Such behavior in this simple system can provide clues in the search for interesting physical behaviors in larger systems, which can be investigated using computer modeling.

\section{Conclusions}

We presented a simple, intuitive way to introduce students to the field of non-equilibrium statistical physics using a one dimensional spin system, the two-temperature kinetic Ising model. We solved exactly the master equation for a 1x4 system and found the probability distribution for both the steady-state and the time dependent case. By comparing the steady-state probability spectrum with the equilibrium counterpart, we were able to see the dramatic differences between the two systems. For example, the non-equilibrium steady state probability distribution is governed not only by the configurational energy (which is the case for equilibrium), but also by other factors, that remain to be explored. The time-dependent probability spectrum offers new features that are worth pursuing (e.g local maxima that may be indicative of an oscillatory behavior of the system before settling into its steady-state). We also emphasized the presence of the probability currents- a defining feature of non- equilibrium systems. 

From a pedagogical point of view, this project offers students the opportunity to observe and learn about the novel behavior of a driven system, and face the challenges of a lack of theoretical framework for far from equilibrium systems. We see this analytical study as a starting point for a more complex project. We plan to study bigger system sizes with the help of Monte Carlo computer simulations. Besides being a useful and necessary tool in the study of various theoretical models, computer simulations are very appealing and accessible to undergraduate students. Programming is an important part of any scientific curriculum. Research projects based on computer simulations give students an opportunity to bridge three core disciplines of their curriculum, physics, computer science and mathematics. Also, computer simulations are a  intuitive way to introduce students to the non-equilibrium statistical physics: using visualization tools, students can "see" how the system goes from an equilibrium to a non-equilibrium state by simply switching off and on the extra heat bath.

\section{Acknowledgments}

We would like to express our gratitude to B. Schmittmann and R.K.P. Zia for their very helpful and encouraging discussions and communications.
This work was supported in part by Thomas F. and Kate Miller Jeffress Grant J-763 and by Robert E. Lee Research Fund for undergraduate research.


\section {Bibliography}

\begin{thebibliography}{99}


\bibitem{Gibbs}  J. W. Gibbs, Elementary Principles in Statistical
Mechanics, (Scribner, N.Y, 1902)

\bibitem{Schmuser}  F. Schmuser and B. Schmittmann\textbf{, }\textit{J.
Phys. A}\textbf{: }\textit{Math. Gen}\textbf{. 35: }2569 \ (2002)

\bibitem{Boltzmann}  L. Boltzmann \textit{Lectures on Gas Theory}, english
translation (Berkeley, California,1964)

\bibitem{Pathria}  R. K Pathria \textit{Statistical Mechanics} 2nd ed
(Butterworth-Heinemann, Oxford, 1996)

\bibitem{zia(paper)} R. K. P. Zia and B. Schmittmann \textbf{, }\textit{J.
Stat. Mech.}\textbf{: }\textit{Theory and Experiment} P07012 \ (2007)

\bibitem{Glauber}  R.J Glauber, \textit{J. Math. Phys}. \textbf{4}:294 (1963)

\bibitem{Lenz}  W. Lenz \textit{Z. Phys }\textbf{56:}778 (1929)


\bibitem{Ising}  E. Ising \ \textit{Z. Phys }\textbf{31:}253 (1925)


\bibitem{RZ(kinetic)}  Z. Racz and R.K.P Zia, \textit{Phys Rev. Lett. E}, 
\textbf{49}:139 (1994)

\bibitem{Mobilia}  M. Mobilia, B. Schmittmann, and R. K. P. Zia, \textit{Phys Rev. Lett. E}, 
\textbf{71}:056129 (2005)


\bibitem{KLS}  S. Katz, J.L Lebowitz and H. Spohn, \textit{Phys. Rev. B} 
\textbf{28}:1655 (1983)


\bibitem{SZ}  B. Schmittmann and R.K.P Zia, \textit{``Phase Transitions and
Critical Phenomena''} Vol 17, edited by C. Domb and J.L Lebowitz (Academic,
London, 1995)


\bibitem{Binder}  K. Binder and D.W Heermann \textit{''Monte Carlo
Simulation in Statistical Physics'' }Springer, Berlin (1988)


\bibitem{Metropolis}  N. Metropolis, A.W. Rosenbluth, M.M Rosenbluth, A.H
Teller and E. Teller, \textit{J. Chem. Phys} \textbf{21}:1097 (1953)





\end{thebibliography}
\end{document}